\documentclass[printer]{aa} 
\usepackage{graphicx}
\usepackage{txfonts}
\usepackage{color}
\usepackage{hyperref}
\def\fdeg{\hbox{$^\circ$}}
\newcommand{\teff}{T_{\mbox{\textrm{\scriptsize eff}}}}
\newcommand{\teffprime}{T^{\prime}_{\mbox{\textrm{\scriptsize eff}}}}
\newcommand{\gk}{G-K_{s}}
\newcommand{\jk}{J-K_{s}}
\newcommand{\ks}{{K}_{s}}
\newcommand{\tnorm}{\hat{T}}

\newcommand{\feh}{\textrm{[Fe/H]}} 
\newcommand{\kG}{\mbox{k}_{G}}
\newcommand{\kK}{\mbox{k}_{\ks}} 
\newcommand{\kJ}{\mbox{k}_{J}}
\begin{document}

\title{3D maps of interstellar dust in the Local Arm: using $Gaia$, 2MASS and APOGEE-DR14}


\author{R. Lallement
\inst{1}
\and
 L. Capitanio\inst{1}
\and
L. Ruiz-Dern\inst{1}
\and
C. Danielski\inst{1,2,3}
\and
C. Babusiaux\inst{4,1}
\and
J.L. Vergely\inst{5}
\and
M.~Elyajouri\inst{1}
\and
F. Arenou\inst{1}
\and 
N. Leclerc\inst{1}}

\institute{GEPI, Observatoire de Paris, PSL University, CNRS,  5 Place Jules Janssen, 92190 Meudon, France 
              \email{rosine.lallement@obspm.fr}
\and 
Irfu/DAp, CEA, Université Paris-Saclay, F-9119 Gif-sur Yvette, France
\and
Université Paris Diderot, AIM, Sorbonne Paris Cité, CEA, CNRS, F-91191 Gif-sur-Yvette, France
  \and
  Univ. Grenoble Alpes, CNRS, IPAG, 38000 Grenoble, France
\and
  ACRI-ST, Sofia-Antipolis, France
}

\date{Received ; accepted }
\titlerunning{Local ISM 3D maps}
 
\abstract
{$Gaia$ data and stellar surveys open the way to the construction of detailed 3D maps of the Galactic interstellar (IS) dust based on the synthesis of star distances and extinctions. Dust maps are tools of broad use, including for $Gaia$-related Milky Way studies.}
{Reliable  extinction measurements require very accurate photometric calibrations. We show the first step of an iterative process linking 3D dust maps and photometric calibrations and improving them simultaneously. 
}
{Our previous 3D map of nearby IS dust was used to select low reddening SDSS/APOGEE-DR14 red giants, and this database served for an empirical effective temperature- and metallicity-dependent photometric calibration in the $Gaia$ \textit{G} and 2MASS \textit{K$_{s}$} bands. This calibration has been combined with $Gaia$ G-band empirical extinction coefficients recently published, \textit{G}, \textit{J} and \textit{K$_{s}$} photometry and APOGEE atmospheric parameters to derive the extinction of a large fraction of the survey targets. 
Distances were estimated independently using isochrones and the magnitude-independent extinction $K_{J-\ks}$. This new dataset has been merged with the one used for the earlier version of dust map. A new Bayesian inversion of distance-extinction pairs has been performed to produce an updated 3D map.}
{
We present several properties of the new map. Its comparison with 2D dust emission reveals that all large dust shells seen in emission at mid- and high-latitude are closer than 300pc. The updated distribution constrains the well debated, X-ray bright \textit{North Polar Spur} to originate beyond 800 pc.  We use the Orion region to illustrate additional details and distant clouds. On the large scale the map reveals a complex structure of the Local Arm. 2 to 3 kpc-long chains of clouds appear in planes tilted by $\simeq$ 15$\fdeg$ with respect to the Galactic plane. A series of cavities oriented along a l$\simeq$60-240$\fdeg$ axis crosses the Arm.}
{The results illustrate the on-going synergy between 3D mapping of IS dust  and stellar calibrations in the context of $Gaia$. Dust maps provide \textit{prior} foregrounds for future calibrations appropriate to different target characteristics or ranges of extinction, allowing in turn to increase extinction data and produce more detailed and extended maps.} 

\maketitle

\section{Introduction}

Three-dimensional (3D) maps of the Galactic interstellar matter (ISM) are a general tool for various purposes: studies of foreground, background or environment for specific Galactic objects, modeling of photon or high energy particle propagation, estimates of integrated Galactic foregrounds, and finally, of particular importance today in the context of $Gaia$, corrections for the reddening of the observed Milky Way stars. The construction of 3D maps requires massive amounts of distance-limited interstellar (IS) absorption data, i.e. they are necessarily based on absorption by IS dust or gas in the light of stellar objects, and the corresponding distances to the targets.

The synthesis of distance-limited absorption data to retrieve 3D information on the IS matter has been done in two main ways. The first method distributes the data into solid angles and treats independently each radial direction.
Following this method, Bayesian derivations of color excess radial profiles have been performed along radial directions, sightline by sightline. The first map was produced by \cite{Arenou92} based on Hipparcos data. Much later, the first map based on a massive survey from ground was done by \cite{Marshall06} who used 2MASS photometry and the Besan{\c c}on model of stellar population synthesis \citep{Robin12} to produce extinction curves at low Galactic latitudes (-10$\leq$b$\leq$+10 $\fdeg$), and, from their derivatives, 3D maps of the IS dust up to 8kpc in the Galactic center hemisphere. \cite{Chen13} then \cite{Schultheis14} used the Besan{\c c}on model and the VVV survey to produce 2D and 3D maps of the Bulge. \cite{Sale14} used photometric data from the IPHAS Survey to reconstruct extinction radial profiles and dust maps at a better resolution for -5$\leq$b$\leq$+5 $\fdeg$ and up to 5 kpc.   \cite{Green15} used 2MASS and PanSTARRS-1 data for 800,000 stars and color-color diagrams to derive reddening radial profiles at very high angular resolution (on the order of 7 $\arcmin$) in the whole region of the sky accessible for these two surveys. In parallel, maps of the Local Interstellar Bubble have been drawn by \cite{Vanloon13} based on neutral sodium and diffuse interstellar band (DIBs) data, and large scale maps of the integrated IS matter based on massive amounts of DIB measurements have been also produced by \cite{Kos14,  Zasowski15}.  

Full 3D inversions differ from the previous technique by imposing spatial correlations between volume densities of IS matter in all directions, therefore linking adjacent sightlines. 3D inversions following a technique developed by \cite{Vergely01} and with mono- then bi-modal covariance functions have been applied to the nearby ISM and much smaller datasets than in the case of the works cited above: columns of gaseous species \citep{Welsh10},  color excesses from ground-based photometry \citep{Vergely10, Lallement14} or composite sources combining photometry and DIBs \citep{Capitanio17}. \cite{SaleMagorrian14,SaleMagorrian15,SaleMagorrian17} developed a Gaussian field method adapted to realistic multi-scale IS matter distributions. \cite{Rezaei17} also developed a 3D inversion method and tested their new technique on extinction-distance measurements derived from the APOKASC catalogue of  \cite{Rodrigues14}. 

The astrometric ESA mission $Gaia$ \citep{Gaia16b} is an unprecedented opportunity to build improved 3D maps of the Galactic ISM. Parallax distances for more than a billion of stars \citep{Gaia16a} will become available for combination with all types of distance-limited absorption data. $Gaia$ will additionally provide extremely precise photometric measurements and in turn new extinction estimates, either from the satellite data only or in combination with ground-based stellar photometric or spectro-photometric surveys \citep{Gaia16b}.  Extinction estimates based on $Gaia$ include wide-band photometry only, i.e. integrated magnitudes in the $G$-band and blue and red photometers (BP and RP) passbands, spectro-photometric determinations additionally based on the two BP and RP spectra, and in future more accurate determinations based on BP, RP, \textit{G} as well as stellar parameters from the Radial Velocity Spectrometer (RVS) spectra are expected. 

RVS measurements will be restricted to bright targets. Prior to the release of the RVS spectra, and in general for all stars too faint to be characterized by the spectrograph,  extinction estimates based on $Gaia$ low resolution data will suffer from the degeneracy between extinction and spectral energy distribution \citep{Liu12}. This is why calibrations of the photometric bands and extinction coefficients as a function of the stellar parameters and the absorption are needed to properly determine the interstellar reddening and exploit $Gaia$ data in conjunction with spectroscopic surveys, either for stellar science purposes or interstellar medium studies.

\begin{figure}
\centering
\includegraphics[width=0.95\hsize]{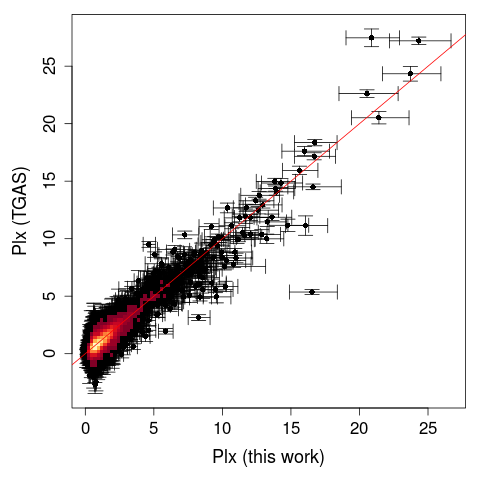}
\caption{Comparison of photometric distances obtained in this work with the $Gaia$/TGAS parallax distances for 13451 APOGEE stars. The colour scale represents the square root of the density of stars, with yellow (resp. black) for the highest (resp. lowest) density.}
\label{tgasdistcomp}
\end{figure}

\begin{figure*}
\includegraphics[width=6cm]{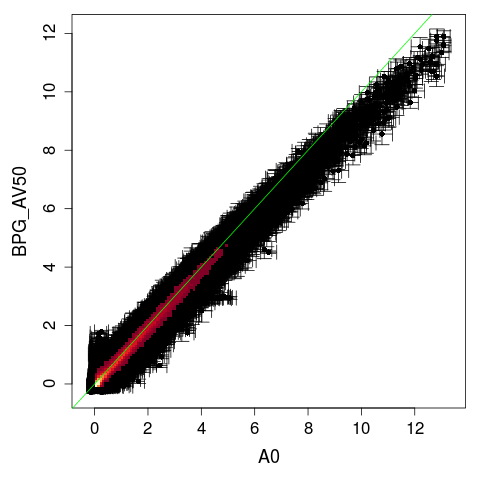}
\includegraphics[width=6cm]{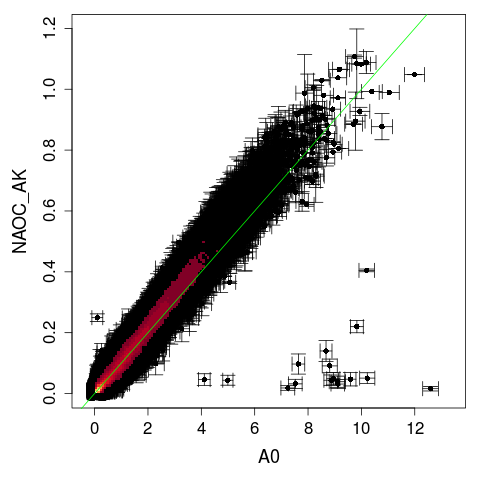}
\includegraphics[width=6cm]{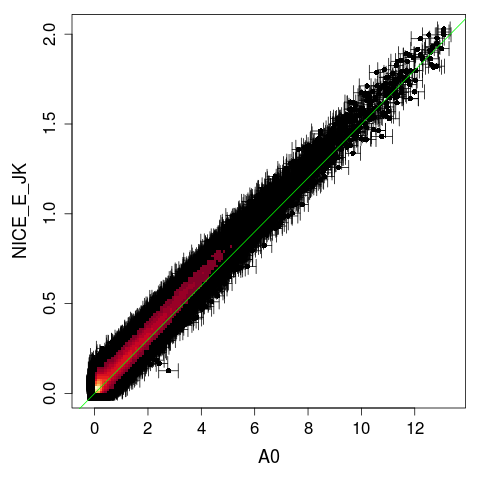}
\caption{Comparison between the extinction $A_0$ derived here with:
-(left) BPG $A_V$ (red line slope : 1.0), -(middle) NAOC $A_K$ (red line slope : 0.1), -(right) NICE $E(J-K)$ (red line slope : 0.15). The colour scale represents the square root of the density of stars, with yellow (resp. black) for the highest (resp. lowest) density.}
\label{fig:A0diff}%
\end{figure*}

3D maps of the nearby ISM may be used to select non or weakly reddened targets to enter empirical photometric calibrations, or more generally as a source of prior solutions for the extinction to be used in further calibrations. Reciprocally, extinction measurements using the new calibrations can feed the distance-extinction databases, allowing the construction of improved and more extended maps. In this work we describe the first step of this iterative approach based on  stellar-interstellar synergy. The 3D maps of \cite{Capitanio17} based on large+narrow band photometry have been used by \cite{RuizDern17} to calibrate the bright TGAS subset of $Gaia$ Red Clump in the $Gaia$ $G$ band, including temperature-metallicity-colour relations using APOGEE-DR13 \citep{Albareti17} stellar parameters. They similarly served to select low reddening targets to enter calculations of effective temperature and absorption-dependent coefficients in the $G$ band \citep{Danielski18}. As part of the present work, the \cite{RuizDern17} calibration has been updated to the APOGEE-DR14 stellar parameters \citep{Abolfathi17}. This extended empirical calibration and the \cite{Danielski18} extinction coefficients have been combined with the APOGEE-DR14 stellar parameters as well as with 2MASS \citep{2mass} and $Gaia$ G band photometric data to derive extinctions for a large fraction of the targets.  Distances to the targets are independently estimated based on isochrones. A new 3D inversion of distance-extinction pairs has subsequently been performed after inclusion of this new dataset in the previous catalog. 

In section 2 we describe the method used to estimate the distances to the APOGEE targets and how the $Gaia$ red giant calibrations performed by \cite{RuizDern17} have been updated to parameters of the APOGEE data release 14. We describe how this calibration and $Gaia$-$G$ band extinction coefficients have been used together with 2MASS and $Gaia$-G band data as well as APOGEE atmospheric parameters to derive individual extinctions for those targets.
In section 3 we describe the resulting additional database and  the new 3D inversion. We compare 2D maps of cumulative reddening up to 300 pc,  800 pc and finally up to the boundaries of the computational volume with dust emission maps. The comparison shows which of the main features seen in emission have been assigned a location. 
We illustrate the dust map improvements, present some new results  as well as new online tools to use the maps in section 4.
In section 5 we describe how the new augmented extinction-distance database has served to add local details to the dust distribution used in the Besan{\c c}on model to enter the $Gaia$ Universe Model Snapshot \citep[GUMS,][]{GUMS_CAT_12}. We conclude and discuss future 3D map improvements in section 6. 

\section{Distance-extinction measurements of APOGEE-DR14 targets}\label{calib_apogee}

For deriving distance and extinction of the APOGEE objects, we 
use photometric data from both $Gaia$ DR1 and 2MASS, and spectroscopic parameters such as effective temperature $\teff$, metallicity $\feh$ and surface gravity $\log(g)$ from APOGEE-DR14 \citep{Abolfathi17,Majewski17}. Here we have used the ASPCAP parameters \citep{ASPCAP}. The cross-match between those catalogues was done using the 2MASS ID provided in APOGEE and the 2MASS-GDR1 cross-matched catalogue of \cite{Marrese17}.

\subsection{Distance}

 We computed the distance modulus of 161683 APOGEE stars by using a Bayesian method on the Padova isochrones \citep[][CMD 2.7]{bressan_isopadova_2012} and the magnitude independent of extinction $K_{J-\ks}$. We adopted the Initial Mass Function (IMF) of \cite{chabrier_galactic_2001} on the mass distribution prior, and selected a flat distribution for the prior on age. A criterion of ${\chi}^2_\textrm{0.99}$ was applied to reject stars too far from the isochrones. 

The results obtained were compared to the four distance estimations provided directly by the APOGEE team\footnote{\url{http://www.sdss.org/dr14/data_access/value-added-catalogs/?vac_id=apogee-dr14-based-distance-estimations/}}: BPG \citep[at 50 percentile,][]{santiago_apogeedistBPG_2016, queiroz_apogeedistBPG_2017}, NAOC \citep{wang_apogeedistNAOC_2016}, NICE \citep{schultheis_apogeedistNICE_2014} and NMSU (Holtzman et al., in prep). They all rely upon the calibrated stellar atmospheric parameters of APOGEE-DR14 and isochrones. The results, as expected, agree, with small differences in particular at the largest distances. The largest differences are with the NICE estimations, with only 0.9\% of outliers at 5 sigma.



We also compared our distances to the $Gaia$ parallaxes for a subset of 13451 APOGEE stars present in the TGAS catalogue (Figure \ref{tgasdistcomp}). We find only 0.1\% of outliers and the normalized residuals follow a normal distribution centered around zero if we account for the small $Gaia$ parallax zero point offset of -0.04~mas \citep{Arenou17}.

\subsection{Extinction}
Similarly to the analysis of \cite{RuizDern17}, made for Red Clump stars, we used the \cite{Capitanio17} 3D maps to select those APOGEE red giant targets for which sightlines are devoid of clouds and extinction is expected to be very weak. We used those targets to update the coefficients of the effective temperature metallicity-dependent calibrations $\tnorm$ vs ($\gk)_0$ and ($\gk)_0$ vs $\tnorm$, where $\tnorm$ = $\teff$/5040 K,  using the last APOGEE-DR14 catalogue. The new results are shown in Table \ref{tab:teffccoeffs}, together with the respective intervals of validity \citep[for the formalism please refer to][]{RuizDern17}.

\begin{table*}[t]
\caption{Updated coefficients and range of applicability of the $\tnorm$ vs. ($\gk)_0$ relation (top table) and the $\tnorm$ vs. ($\gk)_0$  relation (bottom table), $Y = a_0 + a_1\;X + a_2\;X^2 + a_3\;\feh + a_4\;\feh^2 + a_5\;X \;\feh$, where $X$ and $Y$ are either the ($\gk)_0$ colour or the normalised effective temperature $\tnorm = \teff/5040$. The range of temperatures of the [$\tnorm$, ($\gk)_0$] calibration (second table) is given in $\teff$ (not $\tnorm$).} \label{tab:teffccoeffs}
\begin{center}
\resizebox{\textwidth}{!}{\begin{tabular}{cccrrrrrrccr} \hline\hline
\multicolumn{1}{c|}{$\teff$} & ($\gk)_0$  range & $\feh$ range &  \multicolumn{1}{c}{$a_0$} & \multicolumn{1}{c}{$a_1$} & \multicolumn{1}{c}{$a_2$}  & \multicolumn{1}{c}{$a_3$} & \multicolumn{1}{c}{$a_4$}  & \multicolumn{1}{c}{$a_5$} & \multicolumn{1}{c}{RMS$_{[\teff (K)]}$} & \multicolumn{1}{c}{$\%_{\rm outliers}$ } & \multicolumn{1}{c}{N} \\\hline 
$\tnorm$ & [1.6, 3.4] & [-2.3, 0.4] & 1.591 $\pm$ 0.021 & -0.413 $\pm$ 0.018 & 0.048 $\pm$ 0.004 & 0.009 $\pm$ 0.001 & \multicolumn{1}{c}{-} & \multicolumn{1}{c}{-} & 47.7 & 0.2 & 855  \\\hline
\multicolumn{11}{c}{} \\\hline 
\multicolumn{1}{c|}{Colour} & $\teff$ range (K) & $\feh$ range & \multicolumn{1}{c}{$a_0$} & \multicolumn{1}{c}{$a_1$} & \multicolumn{1}{c}{$a_2$}  & \multicolumn{1}{c}{$a_3$} & \multicolumn{1}{c}{$a_4$}  & \multicolumn{1}{c}{$a_5$}   & \multicolumn{1}{c}{RMS$_{[\gk]}$} & \multicolumn{1}{c}{$\%_{\rm outliers}$ } & \multicolumn{1}{c}{N} \\\hline 
($\gk)_0$  & [3761, 5288] & [-2.3, 0.4] & 12.884 $\pm$ 0.423 & -18.592 $\pm$ 0.886 & 7.554 $\pm$ 0.464 & -0.080 $\pm$ 0.115 & 0.022 $\pm$ 0.008 & 0.158 $\pm$ 0.121 & 0.04 & 0.2 & 855 \\\hline
\end{tabular}}
\end{center}
\end{table*}

In a second stage, with the purpose of determining individual extinctions, we selected  from the APOGEE-2MASS-$Gaia$ cross-matched catalogue those stars with effective temperature 3761 K < $\teff \pm \sigma_{\teff}$ < 5288 K
and metallicity -2.3 dex < $\feh$ < 0.42 dex  in order to work within the same limits of the photometric calibration (Tab. \ref{tab:teffccoeffs}). 
We removed all those objects whose temperature error $\sigma_{\teff}$, or metallicity error $\sigma_\feh$ were unknown. 
Finally we discarded stars with photometric errors $\sigma_G$, $\sigma_{\ks}$, $\sigma_J$ > 0.05 mag.
The application of these criteria delivered a sample of  stars.

We used a similar Markov Chain Monte Carlo (MCMC, \citealt{brooks2011}) used in \cite{Danielski18} to properly account for errors.
The MCMC used the jags algorithm \citep{jags} encompassed in \texttt{runjags}
\footnote{\url{https://cran.r-project.org/web/packages/runjags/ runjags.pdf}} library, for R programme language.

For each star in the sample we set effective temperature and metallicity to
follow the normal distribution $\mathcal{N}$  such as
$\teffprime \sim \mathcal{N}(\teff, \sigma_\mathrm{\teff}^2)$ and $\feh^{\prime} \sim \mathcal{N}($\feh$, \sigma_\mathrm{\feh}^2)$, where 
$\teff$ and $\feh$ are the observed temperature and metallicity, and 
$\sigma_\mathrm{Teff}^2$ and $\sigma_\mathrm{\feh}^2$ are the respective observed variances. Consequently $\tnorm^\prime$  was defined as $\tnorm^\prime$ = $\teffprime$/5040.\\
We measured the intrinsic colours ($\gk)_0$ and ($\jk)_0$ using the photometric calibration relations: $(\gk)_\mathrm{0}$  vs $\tnorm^\prime$ (Table \ref{tab:teffccoeffs}) and $(\jk)_\mathrm{0} $ vs $(\gk)_\mathrm{0}$ provided by \cite{RuizDern17}. 

\noindent The likelihood is expressed using star observed colours: ($\gk)_\mathrm{obs}$ and ($\jk)_\mathrm{obs}$ 
\begin{equation*}
\begin{array}{ccl}
(\gk)_\mathrm{obs} &\sim & \mathcal{N}((\gk)_\mathrm{0} + (\mathrm{k}_{G} - \mathrm{k}_{\ks}) \cdot A_0, ~\sigma^2_{G-\ks})\\
(\jk)_\mathrm{obs} & \sim &\mathcal{N}((\jk)_\mathrm{0} + (\mathrm{k}_{J} - \mathrm{k}_{\ks}) \cdot A_0 , ~\sigma^2_{J-\ks})
\end{array}
\end{equation*}

\noindent where $\sigma^2_{G-\ks} = (\sigma_{G}^2 + \sigma_{\ks}^2$) and $\sigma^2_{\jk} = 
(\sigma_{J}^2 + \sigma_{\ks}^2$)
and where $\kG$, $\kJ$ and $\kK$ are the extinction coefficients 
for the $Gaia$ $G$ band and 2MASS $J$ and $\ks$ bands, respectively. These coefficients are expressed as a function of the colour $(\gk)_0$ and extinction $A_0$ as provided by \cite{Danielski18}.
Finally we set the extinction $A_0$ as only free parameter allowed to vary following a uniform prior distribution $\mathcal{U}$ between 0 and 20 mag such as $A_0 \sim$ $\mathcal{U}$(0, 20).

Each MCMC was run using 10$^4$ steps and a burn-in of 4000.
For each star we obtained its extinction mean value and standard deviation. We tested the validity of the results through a $\chi^2_\textrm{0.99}$ criterion and we rejected those stars whose individual extinction did not comply this criterion or whose value was not within the validity ranges of the $\kG$ extinction coefficient. The resulting sample of valid objects consists of 148744 stars.

Extinctions are provided together with distances for the BPG, NAOC and NICE methods, providing respectively $A_V$, $A_K$ and $E(J-K)$. We present in Fig.~\ref{fig:A0diff} the comparison with those studies which globally agrees reasonably well considering the different approaches.

\begin{figure}
\centering
\includegraphics[width=0.95\hsize]{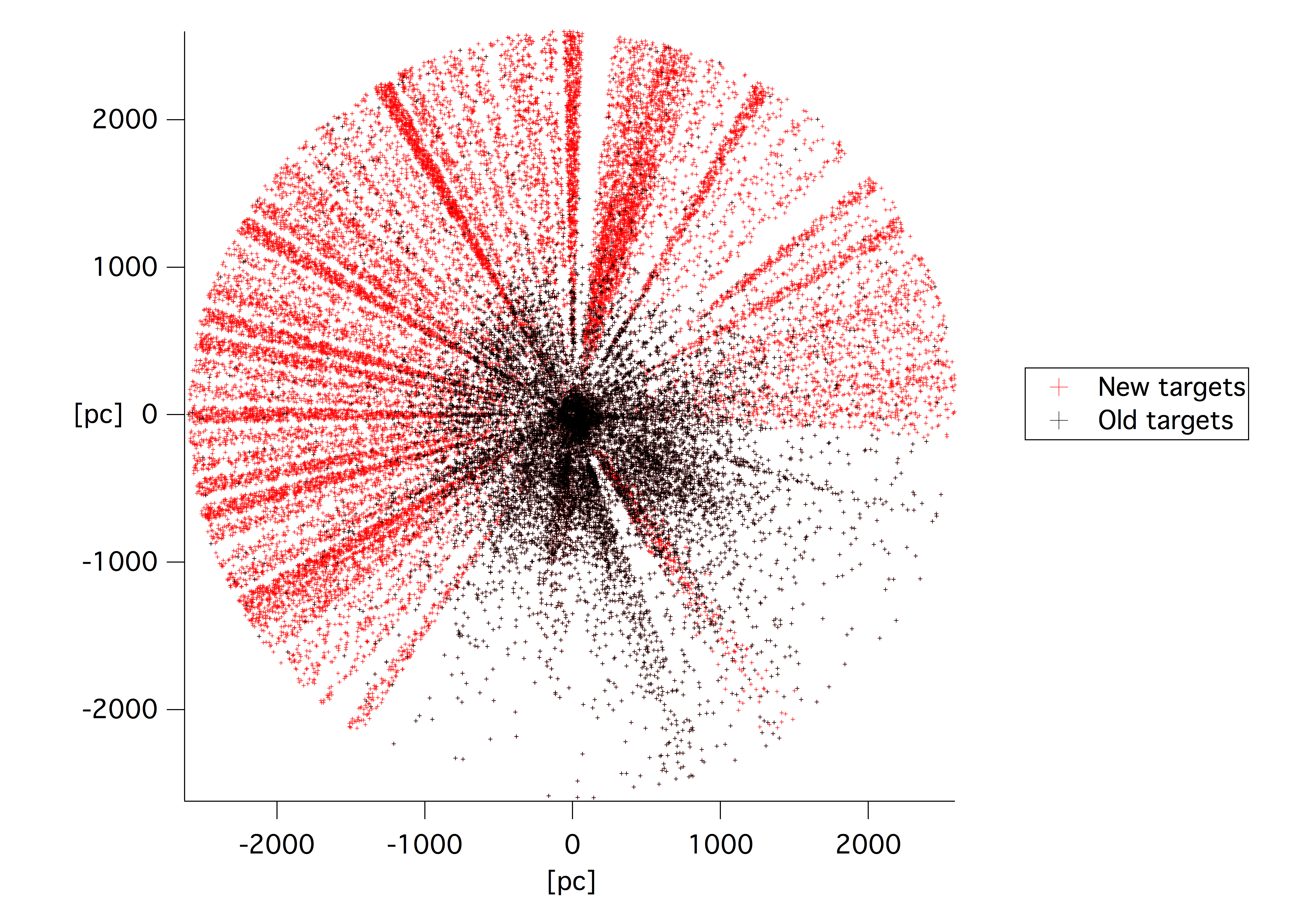}
\caption{Projections onto the Galactic plane of target stars entering the new inversion and with Galactic latitude -10$\leq$b$\leq$+10$\fdeg$. The targets previously used by \cite{Capitanio17} are the black dots. The additional targets are the red dots.}
\label{Fig_targets}%
\end{figure}


\section{Construction of updated 3D maps}

\begin{figure*}[t!]
\sidecaption
\includegraphics[width=0.65\hsize]{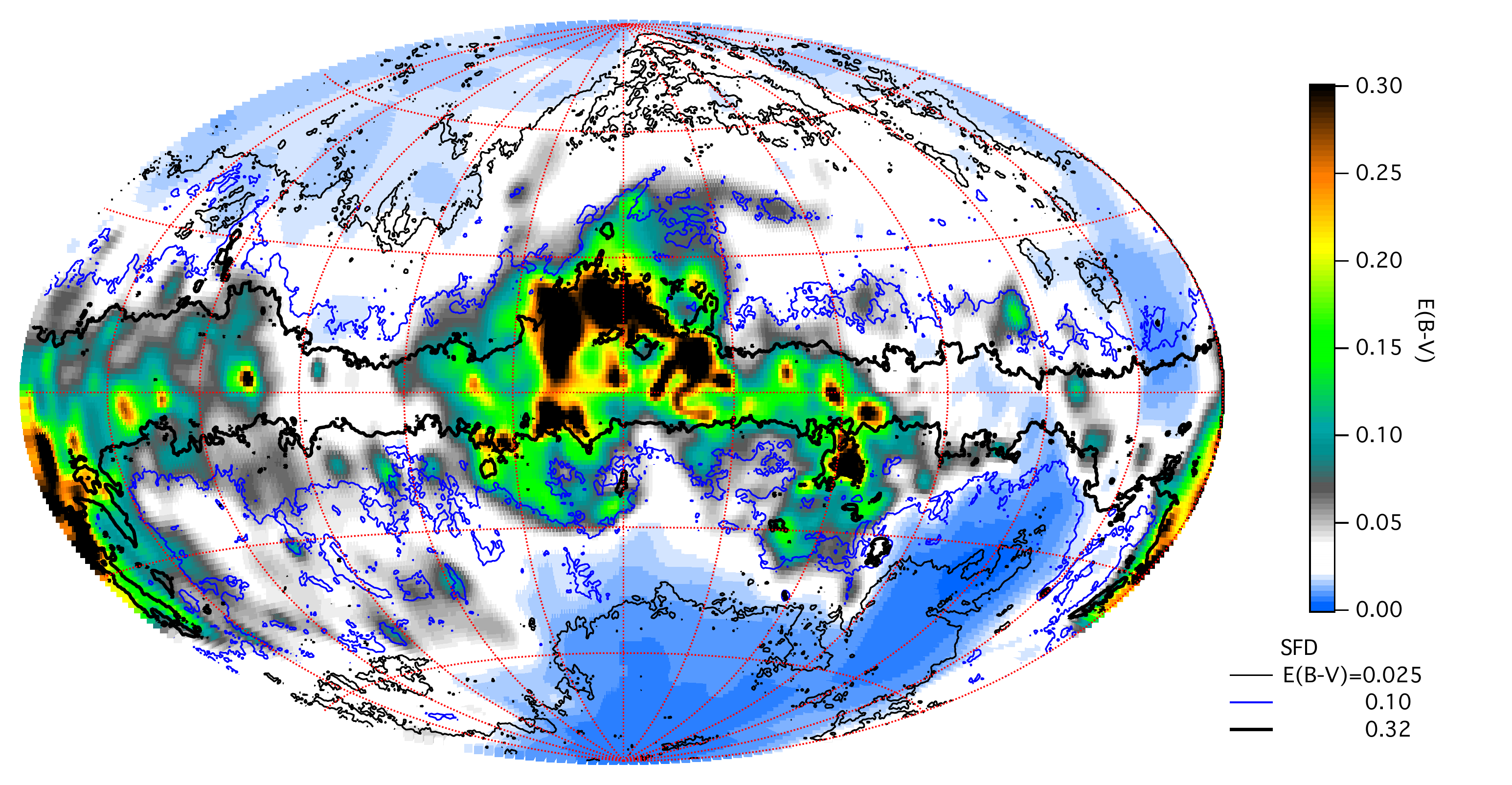}
\caption{Cumulative color excess E(B-V) computed by integration of the differential color excess along radial directions, from the Sun up to a distance of 300 pc. Iso-contours derived from the SFD98 reddening map are superimposed. The comparison shows that all northern high-latitude arches seen at longitudes -150$\leq$l$\leq$+40$\fdeg$ and with E(B-V)$\geq$0.025 mag are closer than 300 pc (see text).}
\label{Figintegcub300}
\end{figure*}

\begin{figure*}[t!]
\sidecaption
\includegraphics[width=0.65\hsize]{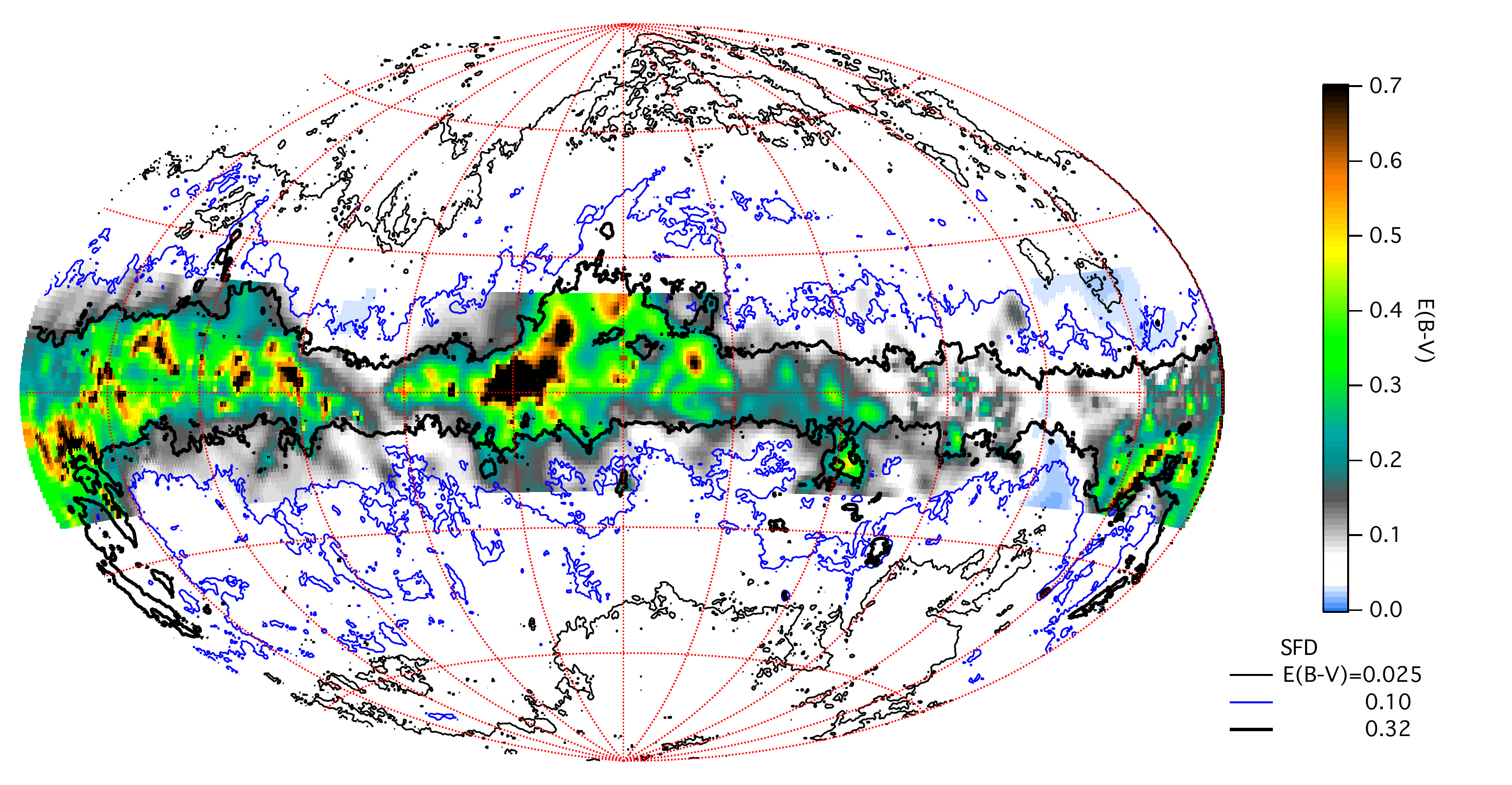}
\caption{Same as Fig \ref{Figintegcub300}, for sightlines 800 pc long. Grid points are left blank at high Galactic latitudes (abs(b)$>\simeq$25$\fdeg$), since for such sightlines 800 pc distant sightline extremities are out of our 4000x4000x600 pc$^{3}$ computational volume. In the second quadrant (90$\leq$l$\leq$180$\fdeg$) there is a good agreement between SFD98 iso-contours for E(B-V)=0.32 mag and locations corresponding to about the same value of our integrated color excess. This shows that most of the structures in these areas are within 800pc. On the contrary, for a large fraction of the third and fourth  quadrants (180$\leq$l$\leq$270$\fdeg$ and 270$\leq$l$\leq$360$\fdeg$ respectively) most of the dust seen in emission is beyond 800 pc.}
\label{Figintegcub800}
\end{figure*}

\begin{figure*}[t!]
\centering
\includegraphics[width=0.9\hsize]{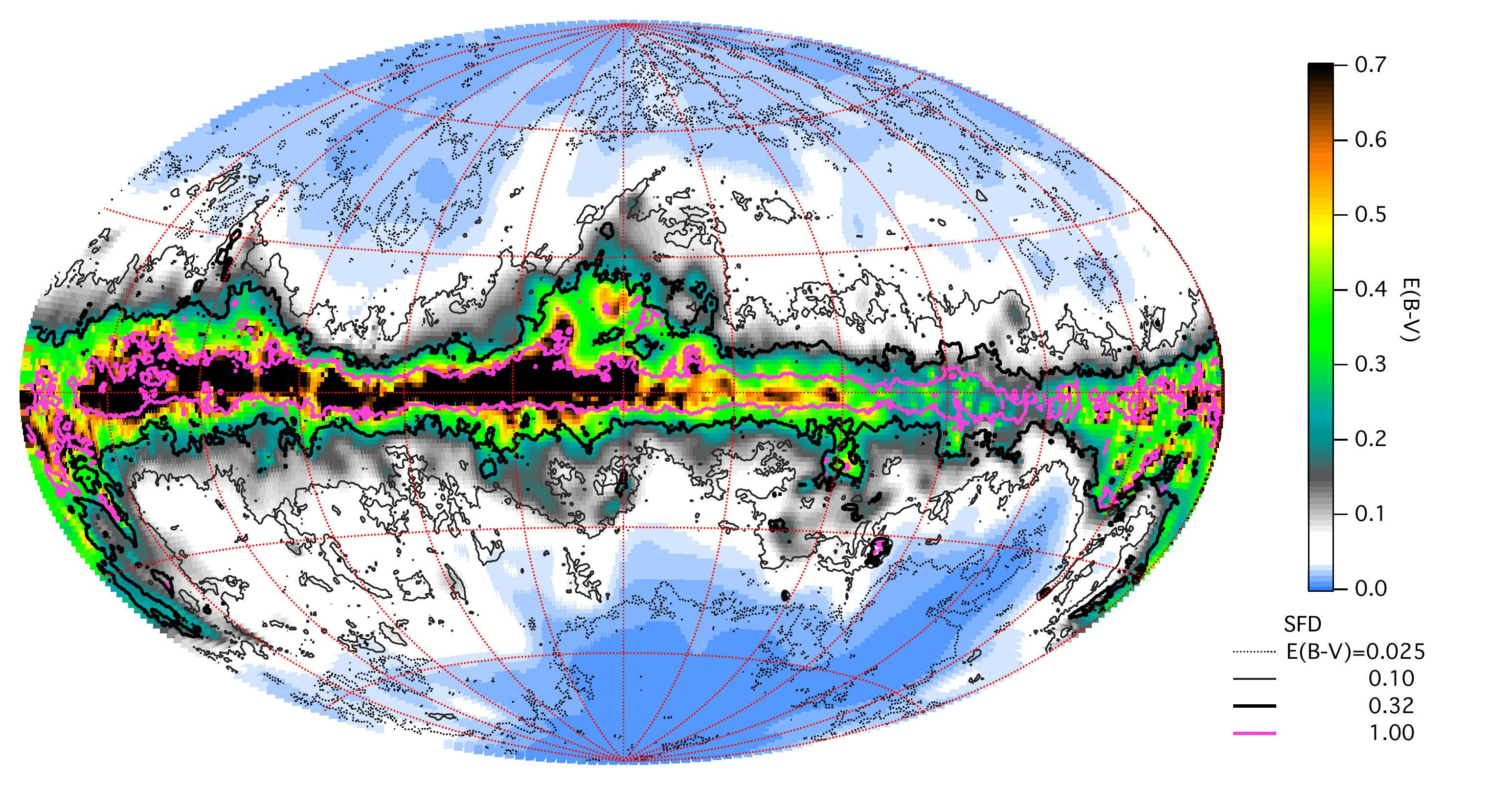}
\caption{Integrated color excess reached at the boundary of the computational volume. Iso-contours derived from the SFD98 reddening map are superimposed.}
\label{Figintegcub}
\end{figure*}

\subsection{Data selection and method}

As described in section \ref{calib_apogee}, distance-extinction pairs and corresponding uncertainties have been obtained for $\simeq$ 149,000  APOGEE-DR14 targets. As in the case of the previously inverted distance-extinction database, we have retained for the inversion all targets for which the estimated distance is smaller than 2.6 kpc, the distance to the Plane Z smaller than 600 pc, and the estimated relative error on the distance smaller than 30\%. 
For the 42633 selected targets following these criteria, extinctions at 550nm A$_0$ were converted into color excess E(B-V) assuming A$_0$ = 3.1 E(B-V). Note that, for this reason, in the following description of the inversion and maps  we use both terms \textit{extinction} or \textit{color excess} since they are simply proportional. 

These new data were merged with those corresponding to the $\sim$30,000 targets used by \cite{Capitanio17}. Doing so, we combined for the first time extinction measurements that were entirely extracted from ground-based photometry and new measurements using both ground- and $Gaia$ space-based G-band photometry. 
We have kept in the former dataset the $\simeq$ 4,900 extinction estimates based on the 15273 $\mu$ diffuse interstellar band extracted from the DR13 APOGEE spectra. As a matter of fact, these estimates were made for APOGEE telluric standard stars (TSSs) and do not overlap with the above targets that are science targets only. The TSSs are brighter and bluer stars and for these objects APOGEE model spectra are not appropriate and the DIBs were extracted following the technique developed by \cite{Elyajouri16}. We also added a few hundred extinctions estimated from uvby$\beta$ photometry by \cite{Kaltcheva14} for targets in the direction of the supershell GSH 305$+$01$-$24  region. Their distance range from 0.1 to 2 kpc is in good agreement with the one of the APOGEE targets (see below). Fig. \ref{Fig_targets} shows the locations of the projection onto the Plane of the initial \cite{Capitanio17} targets and the additional targets, in the two cases for targets with $\vert b \vert \leq$10$\fdeg$. At variance with the \cite{Capitanio17} targets that were located in majority closer than 500 pc, and for a large fraction even closer within the Local Cavity, the new APOGEE science targets are distributed at larger distance and fill the space until 2.0 kpc. This is due to the use of the infra-red spectral domain and the APOGEE sensitivity and illustrates the qualities of the survey. Such a complementarity of distances makes the merged catalog well appropriate for 3D mapping close to the Plane. 

We have performed a regularized Bayesian inversion of all individual color excesses, following the pioneering technique described in \cite{Tarantola82} and successively applied to the local IS dust by \cite{Vergely01}, \cite{Vergely10}, \cite{Lallement14} and \cite{Capitanio17}. 
 This inversion uses sightline-integrated data to create an analytic function of space coordinates that represents the local differential colour excess or differential opacity at any point P $\delta$(E(B-V))/$\delta$d, if d is the distance along any sightline crossing P. This quantity, expressed in mag.pc$^{-1}$, is proportional to the volume density of absorbing dust or volume opacity. The inversion is a (largely) under-constrained problem, and to regularize it we impose a \textit{smoothing} of the volume opacity: the absorbing dust is forced to be distributed in volumes of an imposed minimum size and to have imposed maximum spatial gradients (or contrasts). 
 The correlation function for the opacity in two points is the sum of the two kernels: 
 \begin{equation}
     \psi(x_a,x_b)=\frac{{\sigma_{\chi_0}}^{2}}{\cosh(-\frac{\lvert x_a - x_b \lvert} {\chi_0})}+{\sigma_{\chi_1}}^2 \exp^{\frac{-{\lvert x_a - x_b\lvert}^2}{\chi_1}}
 \end{equation}
 
 In the present inversion the minimum sizes of the structures are $\chi_0 = 30 pc$ and $\chi_1 = 15 pc$, with the corresponding allowed contrasts $\sigma_{\chi_0} = 0.8$ and $\sigma_{\chi_1} = 1.0$. For more details on those choices see \cite{Lallement14}. Uncertainties on the colour excess/extinction and on the target distance both enter the Bayesian adjustment.  As stated above, input data have relative error on distances lower then 33\%. There is no corresponding limitation for the colour excess, however there is a minimum imposed error of 0.01 mag. In case of the DIBs an additional uncertainty is introduced to represent the DIB/extinction variability \citep{Capitanio17}. As detailed in \cite{Lallement14} the inversion algorithm combines the input errors on target distances and color excesses in a single quantity. 

At variance with previous inversions and  for computational reasons, the new 3D inversion was not done using all stellar targets simultaneously. The computation time varies as $ \propto {n_{targets}}^{2}$ if $n_{targets}$ is the number of targets entering the inversion. We chose to compute separately two maps based on targets with longitudes l = (-95 \fdeg{}, 95\fdeg{}) and (85\fdeg{},-85\fdeg{}) respectively, i.e. roughly the Galactic center hemisphere and the Galactic anticenter hemisphere, with an overlap of 5$\fdeg$ on both extremities. 
 The inversion for the "center hemisphere" was done with 35748 stellar targets and the "anticenter" with 38806 ones. In total  71357 sightlines were used from which 3197 were in overlapping regions.
 As in \cite{Capitanio17}, a large-scale prior distribution based on \cite{Green15} has been used for these two inversions. For each inversion we obtained a 3D distribution in a cube $[-2000,2000]pc\times[-2000,2000]pc\times[-300,300]pc$ and this density is discretized in voxels of 5pc side. The merging of the two distributions in the overlapping volume is not a simple average, but instead a weighted average depending on the angular distance to the b= 90$\fdeg$ or b=270$\fdeg$ limiting surfaces. Several checks were performed to ensure that the resulting distribution is very similar to the one one would obtain in a single inversion.

 \subsection{Error estimates}

 The resulting map of volume opacities has intrinsic errors that are due to the regularization, because it tends to smooth the density structures and distribute the opacity in volumes that are wider than the actual ones. Structures smaller than $\sim$ 20 pc like dark clouds or small SNR cavities can not be resolved with the present dataset and algorithm. This introduces also errors on the cumulative color excess computed by integration through the 3D distribution. Uncertainties on cloud locations and opacities are dependent on the target volume density, the clumpiness of the structures as well as on uncertainties on distances and extinctions, i.e., they vary strongly within the computational volume. A precise error calculation for this type of Bayesian inversion method was proposed by \cite{Sale14}, however it is difficult to implement and requires a very long computation time. We decided here to limit our error estimates to the integrated extinction between the Sun and each location in the computational volume. These error estimates can be obtained from the on-line tools described below. An estimate of the uncertainty on the distance to the near side of each structure is also given by the tool.
The distance uncertainty was calculated in each point (x,y,z) of the 3D $[800]\times [800]\times[300]$ voxel matrix (we recall that each voxel center corresponds to 5pc distant points in the $[-2000:2000]$ pc $\times [-2000:2000]$ pc $\times [-300:300]$ pc volume), following the method of \cite{Capitanio17}, i.e., considering that the uncertainty is mainly dependent on the density of targets that constrain the structures. At each point (x,y,z) we computed the target density N in $[200]$ pc $\times$ $[200]$ pc $\times$ $[200]$ pc cubes centered on the point, and the associated distance error was estimated as the inverse of this density: 200 $x$ $N^{-\frac{1}{3}}$.
The error on the cumulative reddening was computed in a different way than in \cite{Capitanio17}. At each point (x,y,z) we used the same set of targets used for the distance uncertainty, i.e. located within a 200 pc  $x$ 200 pc $x$ 200 pc cube around the point, and computed for those targets the mean value of the difference between the measured color excess (the input value) and the one obtained by integration of the final adjusted differential colour excess distribution between the Sun and the target. 
 At large distances where data are scarce this method results in very large errors. Since a limitation  on the color excess is provided by 2D maps of integrated reddening based on dust emission, we used this limit as an upper limit on the integrated reddening between the Sun and any point whatever its distance. Doing so, we allowed for asymmetric errors and computed independently the negative and positive error intervals.
 For the upper limit on the integrated reddening we used the 2D maps of \cite{Schlegel98} (hereafter SFD98). Due to our smoothing constraints the SFD98 map is more appropriate than more recent high resolution maps. At each point we computed the difference between the SFD98 value and the map-integrated reddening and compared this difference with the error calculated following the \textit{neighbooring targets} method previously described. We defined the positive error as the smallest of these two numbers. As a result of this choice, in some very rare cases where our integrated colour excess is higher then the SFD98 value, the positive error becomes null. On the other hand, because the cumulative reddening must be positive, the negative error is chosen as the smallest of the two quantities: our computed error or the reddening itself.
 
 
\subsection{Inversion results: cumulative reddening maps}

Figures \ref{Figintegcub300}, \ref{Figintegcub800} and \ref{Figintegcub} show 2D maps of the cumulative color excess from the Sun to 300 pc, 800 pc and the boundaries of the computational volume respectively. These cumulative color excess maps were simply computed by integrating the local differential color excess resulting from the inversion up to each chosen distance, for a grid of radial directions separated by 1$\fdeg$ in Galactic longitude and latitude. Superimposed are a few selected iso-contours of the total Galactic reddening computed by \cite{Schlegel98}. The comparisons between the maps and the total SFD98 reddening do not allow to estimate errors in our mapping, instead they allow to disentangle structures that have been assigned a location in the 3D map within the chosen distance from structures that are located farther away. 

The 300 pc-integrated maps are of particular interest. In the Northern hemisphere, they clearly show strong coincidences between the SFD98 iso-contour at E(B-V)= 0.025 mag and the directions where the cumulative reddening reaches this value (see the colors and color coding in Fig \ref{Figintegcub300}). This implies that all of the northern high latitude structures seen in the maps and characterized by E(B-V) values on the order of few hundreds of a magnitude are closer than 300 pc. This was already perceptible in our previous maps (see Fig 11 from \cite{Lallement14}) but is seen in a much clearer way here thanks to the addition of high latitude APOGEE targets. 
On the contrary, in the Southern hemisphere the same SFD98 iso-contour at 0.025 mag falls in directions for which the 300pc integrated reddening is weaker (darker blue color), suggesting that either a significant fraction of the emitting dust is beyond this distance or the  dust properties are different. This is particularly marked at longitudes centered around l=240$\fdeg$, i.e. towards the conspicuous Canis Major cavity below the Plane that is already known for containing a large quantity of ionized gas and for its low dust to gas ratio \citep{Lallement15}.

\begin{figure*}[t!]
\centering
\includegraphics[width=0.8\hsize]{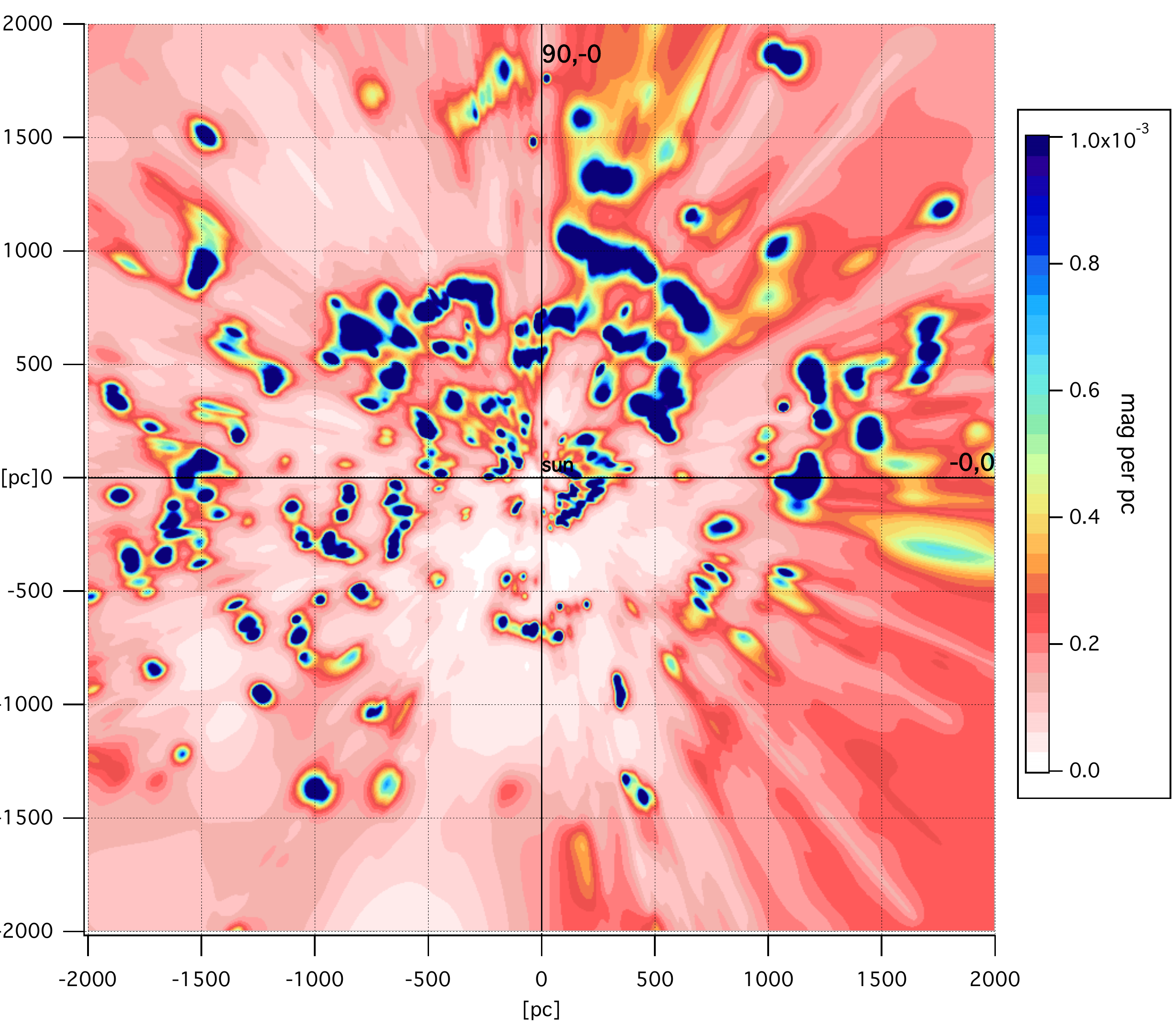}
\caption{Differential color excess in the Galactic Plane computed as a planar cut along the Galactic equator through the 3D distribution. The Sun is at the center of the image (coordinates 0,0). The Galactic centre direction is to the right. The units are mag.pc$^{-1}$. The map can be directly compared with the previous similar map from \cite{Capitanio17}.}
\label{Fig_galplane}%
\end{figure*}


\begin{figure*}
\sidecaption
\includegraphics[width=0.6\hsize]{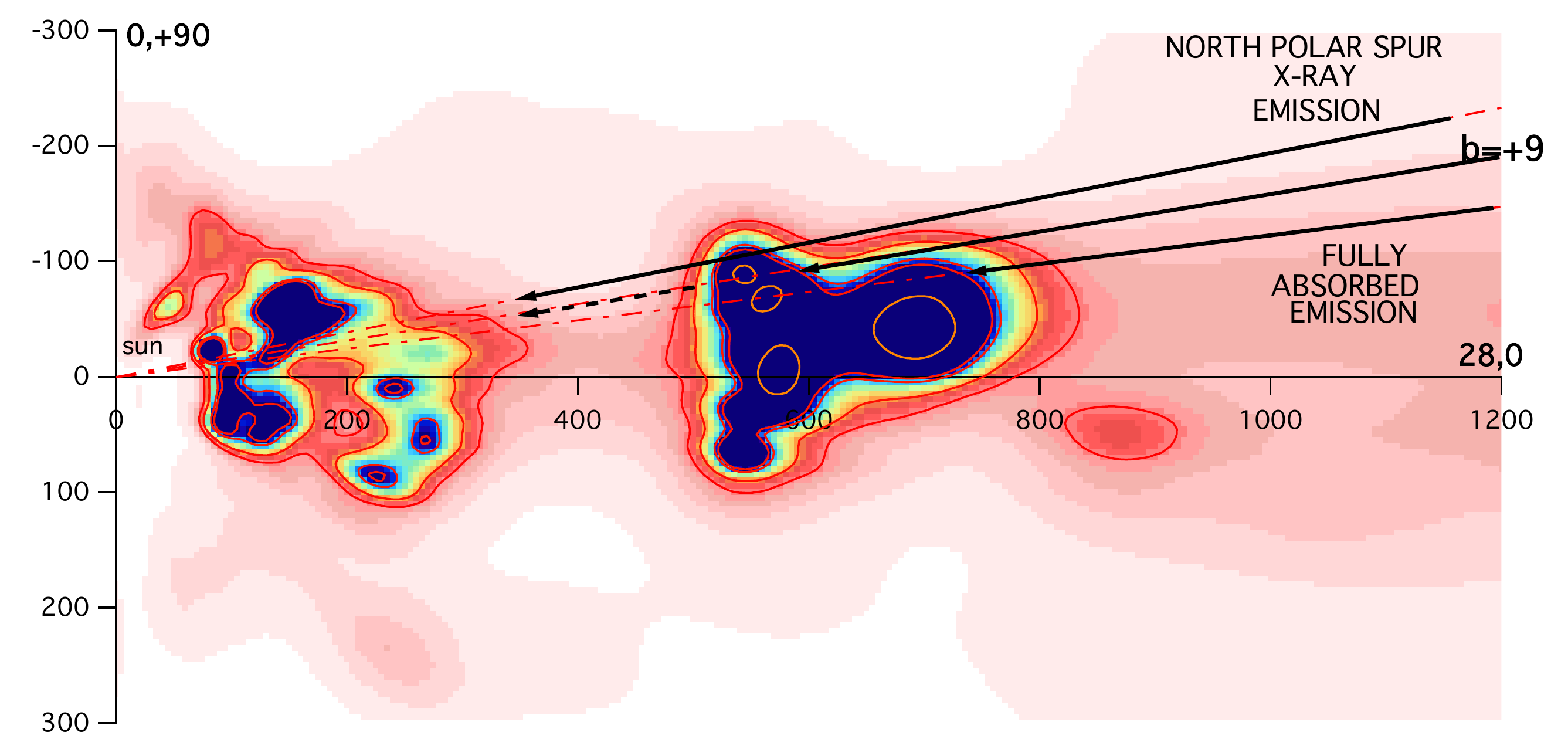}
\caption{Differential color excess in a vertical Plane containing the Sun and along the Galactic longitude l=28$\fdeg$. The North Galactic pole is to the top. The map shows the distribution of Aquila Rift clouds and more distant structures. The b=+9$\fdeg$ dashed line corresponds to the latitude above which X-rays from the North Polar Spur (NPS) start to get detected. The  very thick cloud absorbing predominantly the NPS at b$\leq$+9$\fdeg$  is located between 500 and 800 pc.}
\label{Fig_NPS}%
\end{figure*}

\begin{figure*}
\includegraphics[width=0.98\hsize]{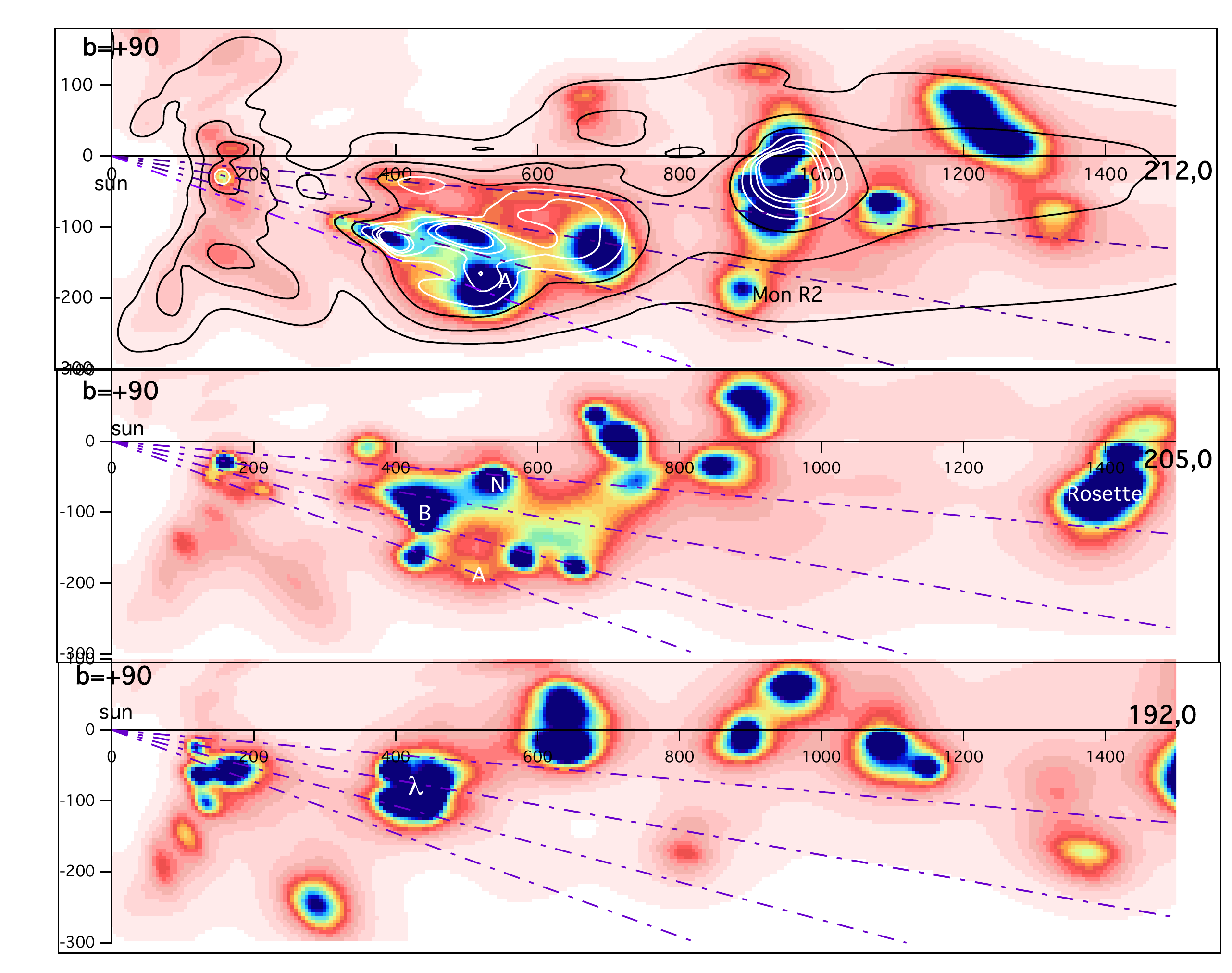}
\caption{Differential color excess in a vertical Plane containing the Sun and along the Galactic longitudes l=212, 205, and 192 $\fdeg$ from top to bottom. The maps show the distribution of the main Orion clouds. Superimposed as black or white lines on the top graph are iso-contours of the differential color excess from the previous inversion of \cite{Capitanio17}. The new 3D distribution reveals more clouds at large distance (e.g. the clouds beyond 800 pc) and resolves more structures in the main Orion region between 400 and 700 pc. We indicate the structures A,B,N as well as $\lambda$ Ori, Mon R2 and Rosette at the locations predicted by \cite{Lombardi11,Schlafly15}.}
\label{Fig_Orion}%
\end{figure*}

\section{3D dust maps and tools}

\subsection{Online tools}\label{Tools}

The website dedicated to the use of the 3D maps \footnote{\url{http://stilism.obspm.fr}} has been updated. Three types of requests can be submitted: (i) Maps of the differential reddening, i.e. equivalently maps of the dust clouds, can be obtained for any planar cut through the computational volume. Differential reddening iso-contours are added to help visualize the faint structures and cloud centers. A requested planar map is defined by the spherical coordinates of its center and  the Galactic coordinates of the normal to the Plane. Zooming in the image is possible and images can be downloaded.  Several examples of use of this tool are given below. (ii) The entire 3D distribution can be downloaded. This may be useful to improve computations of energetic particle propagation, of radiation field or for comparisons with evolutionary models of the ISM. (iii) Cumulative reddening curves with associated estimated uncertainties on the reddening and on the distances to the reddening jumps can be drawn on request for any direction and downloaded. This allows to estimate the reddening at a given distance in a given direction. 

Regarding those tools, we emphasize again that the main utility of our inverted maps is to provide the 3D localization and shape of the main nearby structures, i.e. the information contained in the 2D or 3D images representing the distribution (tools (i) and (ii)). The cumulative reddening curves  (tool (iii)) provide the distances where reddening jumps occur, i.e again information on the distances to the structures and their shapes, however we caution that the absolute value of the reddening may be uncertain in clumpy regions due to the imposed redistribution of the absorbing matter in clouds of sizes larger than 15pc. In such regions of the sky,  tools based on more resolved maps (e.g. the \cite{Green15} map) provide more reliable quantitative estimates.

\subsection{Dust distribution in various planes}
Figure \ref{Fig_galplane}  displays the differential color-excess in the Galactic Plane. 
The comparison with the map based on our previous inversion (Fig. 5 from \cite{Capitanio17}) demonstrates that the addition of the APOGEE targets has resulted in significant mapping extension and in additional details. On the large scale, interestingly the new  structures that are now located beyond 1 kpc in the first quadrant at longitudes comprised between l=0 and l=45 $\fdeg$ appear as some prolongation of the series of 500-1000 pc distant clouds in the fourth quadrant. Together, all those clouds seem to mark the forefront of the Sagittarius Arm. However, based on distributions outside the Plane, as will be discussed below, the boundary between the Local Arm and Carina-Sagittarius in the first quadrant is not as clearly defined as it may appear here in the Galactic plane.

Mapping improvements and extensions may be  illustrated using cuts in the 3D distribution along vertical planes, and we show here two regions of particular interest. Fig \ref{Fig_NPS} corresponds to the vertical plane along the longitude l=28$\fdeg$. This region has been investigated several times due to the existence of the conspicuous X-ray feature called the North Polar Spur (NPS) that brightens above b=+9$\fdeg$ in the 20-30$\fdeg$ longitude range and whose origin is still a matter of lively debate \citep{Planck2015_Pol,Sun2015, Sofue2016}. According to most authors the X-ray emission originates very close to the Sun at $\sim100$ pc in a nearby hot gas cavity blown by one or more supernovae, but its link with the Galactic Center activity and more recently with the Fermi bubbles seen in gamma rays has been also claimed (see \cite{Lall16_NPS} for the various assumptions and references). 3D maps constitute a useful tool to locate the NPS source location. \cite{Puspitarini14} searched for the potential cavity responsible for the X-rays but could not find any nearby wide cavity that could be identified with the NPS. More recently \cite{Lall16_NPS} derived from dedicated XMM-Newton data that the NPS is absorption-bounded and not emission-bounded. The authors used 3D local and large-scale maps from 2MASS and PanSTARRS-1 \citep{Green15} to demonstrate that the NPS emission is absorbed by an opaque cloud  located beyond at least 300 pc, and they found evidence for a much more distant origin. Fig \ref{Fig_NPS}  now reveals very clearly the massive absorbing structure that blocks the X-rays below $\sim$ b=9$\fdeg$ and shows that this absorbing cloud complex is wide and that its far side is at $\sim$800 pc. This provides new and strong evidence that the NPS source is beyond 800 pc, i.e. definitely outside the Local Arm. We believe that there is a fortuitous spatial coincidence between the extended northern HI \textit{arches} seen with 21cm data and the wide and conspicuous radio/X-ray NPS/Loop 1 structure extending to b=+75$\fdeg$. The former are local, as demonstrated by \cite{Puspitarini12} from absorption data for one of the shells and by the 300 pc-integrated emission map of Fig \ref{Figintegcub300}  discussed in the previous section. The latter is generated much farther away.

Fig \ref{Fig_Orion} illustrates the effects of the additional APOGEE targets in the region of the well known Orion clouds. The top graph displays the new inverted dust distribution in a vertical plane along the longitude l=212$\fdeg$. Superimposed are iso-contours of the differential color excess from the \cite{Capitanio17} distribution, i.e. without APOGEE targets. The comparison clearly reveals the addition of more distant clouds in the new maps, especially beyond 500 pc. At short distance there are no noticeable changes, however in the intermediate distance range (say, between 400 and 600 pc), the Orion clouds display more sub-structures. Together, the three graphs in Fig \ref{Fig_Orion} show how the cloud distribution evolves with longitude, and reveal a very complex spatial distribution. We have compared these results with those from \cite{Schlafly15} that are based on PanSTARRS photometry and found full compatibility with the range of distances presented by the authors for the most important structures in their Fig. 2 and 4. We also tentatively located in Fig. \ref{Fig_Orion} the structures identified by \cite{Lombardi11} as A,B,N, as well as $\lambda$Ori, Rosette and Mon R2. For all of them there are clouds at distances and directions in very close agreement with those from \cite{Lombardi11}.

\begin{figure*}
\centering
\includegraphics[width=0.9\hsize]{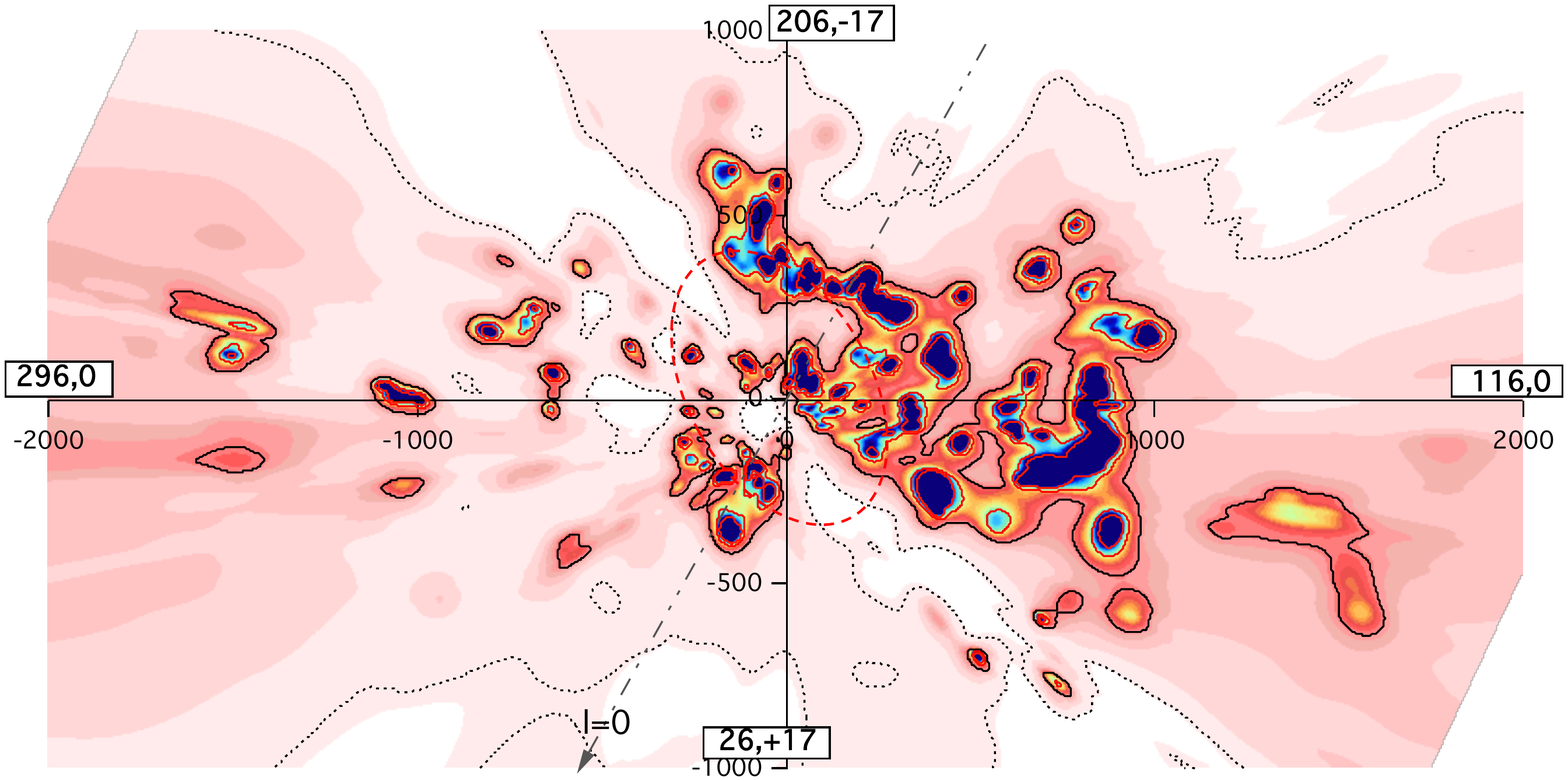}
\caption{Planar cut in the 3D distribution of differential color excess for an inclined plane that does not contain the Sun. The image is centered on the location defined by the spherical coordinates  (l,b,d)= (180.4$\fdeg$, 0$\fdeg$, 104pc). The plane is inclined by 17.4$\fdeg$ w.r.t. the Plane and the longitude of the ascending node is l=296.1$\fdeg$. It corresponds to the Gould Belt plane defined by \cite{Perrot03}. The units are mag. pc$^{-1}$. The coordinates of the four axes are indicated in white bowes. The ellipse found by \cite{Perrot03} to best represent Gould belt clouds and motions as an expanding and rotating structure is superimposed (dashed red line). Note that for this plane the boundary of our computational volume is reached at about 1 kpc along the Y axis.}
\label{Fig_Gould}%
\end{figure*}

\begin{figure}
\centering
\includegraphics[width=0.99\hsize]{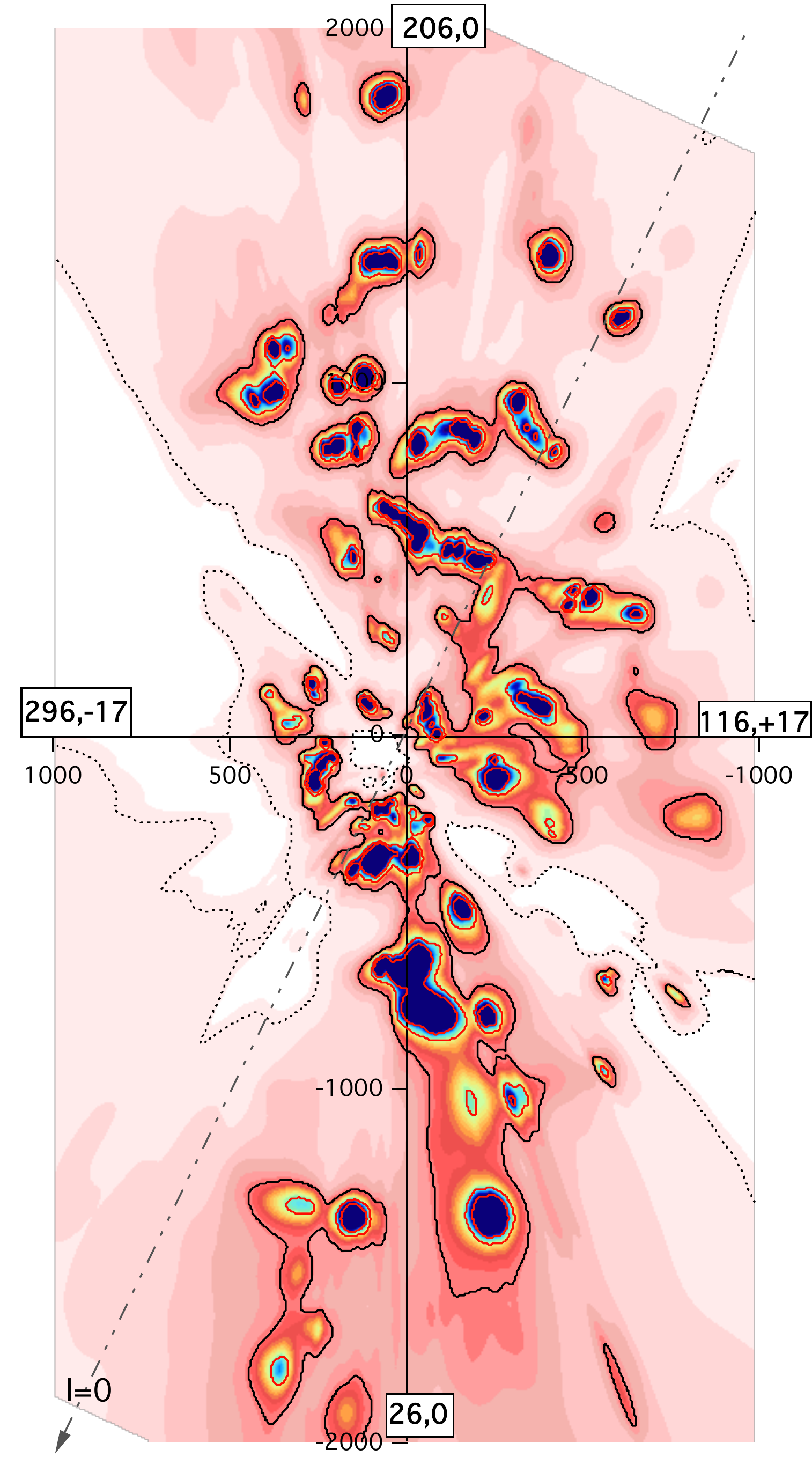}
\caption{Same as fig \ref{Fig_Gould} after rotation of the plane by 90$\fdeg$ around a vertical axis. The inclination with respect to the Galactic plane and the center of the image are the same. The coordinates of the four axes are indicated in white boxes. Note that along the X axis the boundary of our computational volume is reached at about 1 kpc.}
\label{Fig_bibi}%
\end{figure}

As an illustration of a plane that is not Sun-centered and has any orientation, we show in Fig. \ref{Fig_Gould} the cloud distribution in the Gould belt plane as defined by \cite{Perrot03}. The center of the image is 104 pc distant from the Sun along the direction (l,b)=(180.4$\fdeg$, 0$\fdeg$). The plane is inclined by 17.4 $\fdeg$ with respect to the Galactic plane and the ascending node is at the longitude l=296$\fdeg$. It contains the Aquila rift clouds above the Plane on the Galactic center hemisphere side, and Taurus and the chain of Orion clouds below the Plane on the anti-center hemisphere side. We have added the ellipse computed by \cite{Perrot03} as the closest solution for an expanding and rotating structure adjusted to the locations and motions of the main Gould belt structures. The figure strongly suggests that instead of a ring of clouds, the Gould belt is made of tho distinct regions separated by a series of cavities and running parallel to this series (see \cite{Lallement15} for a recent discussion of the Gould belt structure and origin). Fig \ref{Fig_bibi} is similar to Fig. \ref{Fig_Gould}, the unique difference being the longitude of the ascending node, now at l=26$\fdeg$, i.e., this plane can be obtained by rotating the previous Gould belt plane by 90$\fdeg$ along a vertical (b=90$\fdeg$)  axis. The image in this plane shows clouds distributed along $\sim$4 kpc along the l$\simeq$35-225 $\fdeg$ direction, far from the expected orientation of a Galactic Arm approximately aligned with the direction of rotation. this orientation is in better agreement with a \textit{bridge} between Carina-Sagittarius and Perseus. For this reason, the name \textit{Orion bridge} sometimes employed for the Local Arm seems more appropriate. Again, in this plane there seems to be a splitting of the whole structure  into two main regions separated by a series of cavities, including the 150 pc wide Local Cavity in the middle, and two very long and wide cavities towards longitudes l=$\sim$60$\fdeg$ and $\sim$240$\fdeg$. It is now well established that  the Local Bubble is filled with soft X-ray emitting  hot gas, closing years of controversies about its contribution to the soft X-rays diffuse background \citep{Izmodenov99_Sirius, Puspitarini14, Galeazzi14,Snowden15}, and the wide cavity in the third quadrant does also contain hot, X-ray emitting gas \citep{Puspitarini14}. However, it is unclear whether the elongated cavity seen in Fig \ref{Fig_bibi} at  l=$\sim$60-70$\fdeg$  is also filled with hot gas. Further maps, measurements of the physical state of the IS gas in all cavities and of the cloud motions should help shedding light on the source of the peculiar alignment seen in Fig. \ref{Fig_galplane}, \ref{Fig_Gould} and \ref{Fig_bibi}. 

%

%


\section{Integration of local maps in the GUMS}

 3D dust maps of the entire Milky Way are a major ingredient of the Besan{\c c}on model \citep{Robin03}  and the $Gaia$ Universe Model Snasphot \citep[GUMS,][]{Robin12,GUMS_CAT_12} which simulates the entire sky as seen by $Gaia$. As a matter of fact, GUMS computes the extinction created by an imposed 3D distribution for each of all simulated stars of the Besan{\c c}on model. Previous versions of GUMS were based on the 3D dust model of \cite{Drimmel03}, and 
 an updated version of the Besan{\c c}on model\footnote{http://model2016.obs-besancon.fr/modele\_descrip.php} allows to include the \cite{Marshall06} at low latitudes  (-10$\leq$b$\leq$+10$\fdeg$). 
 
 As mentioned in the introduction, the \cite{Marshall06} method used the Besan{\c c}on model of stellar populations and adjusted a 3D distribution of dust to reproduce the counting statistics on magnitudes and colors from the 2MASS survey. 
  Because it uses star counting and the massive 2MASS survey, the \cite{Marshall06} method allows a good representation of the extinction in distant and opaque regions. At variance with the \cite{Marshall06} maps, our local maps described above that are based on measured absorptions of the light of individual stars have their maximum amount of details within the first few hundred parsecs, and they cover the whole sky. However, due to strong biases in target visibility that are linked to the extinction, they do not reproduce well the most opaque clouds. 
The two types of information are, to a certain way, complementary, and for this reason 
we have started to  couple them, i.e. the catalog of individual extinctions presented above and the \cite{Marshall06} model.   Note that, at least for this first attempt, we did not aim at achieving the same spatial resolution as in the above inverted maps, but only a more modest one of 100 pc. 
 The resulting combined model will be used in future GUMS simulations, and comparisons will be made with the previous results, here we only describe the method used to merge the two informations, and present it as an additional illustration of the ongoing stellar-interstellar synergy in the context of $Gaia$. 
 
 
 
For such a merging, the Bayesian approach is particularly well adapted. The \cite{Marshall06} model was taken as the \textit{prior} distribution close to the Galactic plane ($\pm$ 10 $\fdeg$ in latitude), while at higher latitudes we used as  \textit{prior} a dust density that decreases exponentially with the distance to the Plane, has a scale height of 200 pc, and is otherwise invariant. Using a scale height above average measurements avoids the loss of tenuous high latitude structures during the inversion, which may happen when the prior is too low. The information contained in the individual extinctions of the nearby stars was then used to constrain the local distribution, following the formalism described in \cite{Vergely10} and after replacement at low latitudes of the galactic exponential \textit{a priori} model with the Marshall model. Because the latter model is available under the form of extinctions at increasing distance intervals for series of radial directions that are independent of each other, we have maintained this approach, i.e. we have treated each direction independently and selected, around each of them, those observed targets that fall the closest to the chosen sightline.  
More specifically, the 3D space is decomposed in cubic cells of 50 pc size and for each cell we  computed the average of the extinction of all target stars it contains. As a result of this choice, sub-structures smaller than 100 pc are not solved by this model, as noted above. The standard deviation of the extinction of stars in each cell is used to estimate the variability of the extinction inside the cell and this variability is considered as the error on the extinction. A limitation is made according to the number of targets falling in the cell: when this number is below 3, the cell is not used. For each direction, we assembled all validated cells crossed by the sightline, and a Bayesian optimization of the cumulative extinction was performed up to d=1kpc or Z=600pc, whichever was reached first.  Beyond this limit the extinction is constrained to increase linearly with distance until it reaches the one of the Marshall model at 1,5 kpc. Beyond 1.5 kpc the Marshall model is used. The new GUMS model including the local maps and other improvements will be presented elsewhere by the Besan{\c c}on group.

\section{Summary and perspectives}

In this work we have illustrated the on-going synergy between stellar calibrations based on $Gaia$ and ground-based surveys on the one hand, and the construction of 3D maps of the IS dust on the other hand. 3D maps of the local ISM based on wide+narrow band photometric determinations of color excesses of individual target stars and TGAS or photometric distances have been used to select non or weakly reddened targets. These selected targets have been employed to obtain effective temperature-metallicity dependent calibrations of the $(G-\ks)_0$ color for APOGEE red giants. The latter calibrations and subsequent computations of color- and absorption-dependent extinction coefficients in the $G$ band have been used to derive extinction values for most of APOGEE-DR14 red giant targets. Distances were calculated in parallel based on Padova isochrones and the magnitude independent of extinction $K_{J-\ks}$.

The new distance-extinction data were added to the database used for the initial 3D dust maps. Doing so, we associated for the fist time extinction estimates from ground photometry only with estimates from both ground (2MASS, SDSS/APOGEE) and space ($Gaia$) data. A Bayesian inversion of the combined, more than $\sim$twice larger extinction catalog was performed. The resulting 3D maps of dust contain a large number of additional structures, especially at distances beyond 400-500 pc. To illustrate the improvements, we showed 2D maps of cumulative  reddening integrated up to 300, 800 pc, and finally up to the boundary of the 4000x4000x600 pc$^{3}$ computational volume and we compared with total reddening maps deduced from emission data. We showed the dust cloud distribution in several planes. (i) All large scale dust features above $\simeq$ b=+20$\fdeg$ or below b=-20$\fdeg$ were found in the 300 pc integrated maps, showing they are local. (ii) In the Galactic plane the addition of more distant clouds starts to reveal a gap between the Local Arm and the Carina-Sagittarius Arm. (iii) In the Aquila rift region, mapping improvements allow to bring more stringent constraints on the minimum distance to the source of the X-ray bright North Polar Spur (NPS), now found to originate beyond 800 pc. (iv) The spatial distribution of the clouds in the Orion region is now much better defined and we could identify in the 3D maps the structures found by \cite{Lombardi11} and \cite{Schlafly15}. (v) In the Gould belt plane inclined  with respect to the Galactic plane, the map reveals two distinct cloud complexes separated by a series of cavities, instead of an ellipsoidal structure. (vi) In a plane similarly inclined by 17$\fdeg$ but with a 90$\fdeg$ different orientation, the Local Arm appears to be made of a $\geq$3kpc long chain of structures oriented along a  l$\simeq$ 35-225 $\fdeg$ direction, again with a dividing line made of cavities oriented along a l$\simeq$ 60-240 $\fdeg$ axis. The Local Arm in this plane looks like a double \textit{bridge} between Carina-Sagittarius and Perseus. 

New on-line tools were made available to use the 3D distribution of dust (\url{http://stilism.obspm.fr}). It is possible to draw an image of the dust distribution in any chosen plane, containing or not the Sun. We emphasize that this type of tomographic representation is well adapted to full 3D inversions, because such inversions have the advantage of showing how cloud complexes are related to each other in 3D space. As we already warned in previous similar works, absolute values of the integrated extinction between the Sun and each point in the computation volume have uncertainties due to the regularization constraints, in particular they may underestimate the extinction in the cases of sightlines crossing cloud cores. This loss of angular resolution is the price to pay for using full 3D correlations between neighboring points. All images as well as cumulative extinctions can be downloaded, as well as the 3D distribution itself.

The productive feedback between 3D dust maps and photometric calibrations we have illustrated here may be further developed, especially in the context of future $Gaia$ data releases. Once  3D maps reliable up to a given distance are obtained, they can be used a prior solutions for the derivation of extinctions towards more distant stars and for further calibrations. In parallel, $Gaia$ parallaxes with increasing accuracy will become available during the whole mission, allowing to refine photometric determinations of the extinction and to constrain more tightly the distances to the absorbing clouds.

%
   
%
%

\begin{acknowledgements}

R.L.,  JL. V., C.B., F.A. acknowledge support from "Agence Nationale de la Recherche" through the STILISM project (ANR-12-BS05-0016-02).
L.C. acknowledges doctoral grant funding from the Centre National d'Etudes Spatiales (CNES). 
M.E. acknowledges funding from the "Region Ile-de-France" through the DIM-ACAV project. C.D. acknowledges the post-doctoral grant funding from Centre National d’etudes Spatiales (CNES) and support from the LabEx P2IO, the French ANR contract 05-BLAN-NT09-573739.

Funding for the Sloan Digital Sky Survey IV has been provided by the Alfred P. Sloan Foundation, the U.S. Department of Energy Office of Science, and the Participating Institutions. SDSS-IV acknowledges
support and resources from the Center for High-Performance Computing at
the University of Utah. The SDSS web site is www.sdss.org.

SDSS-IV is managed by the Astrophysical Research Consortium for the 
Participating Institutions of the SDSS Collaboration including the 
Brazilian Participation Group, the Carnegie Institution for Science, 
Carnegie Mellon University, the Chilean Participation Group, the French Participation Group, Harvard-Smithsonian Center for Astrophysics, 
Instituto de Astrof\'isica de Canarias, The Johns Hopkins University, 
Kavli Institute for the Physics and Mathematics of the Universe (IPMU) / 
University of Tokyo, Lawrence Berkeley National Laboratory, 
Leibniz Institut f\"ur Astrophysik Potsdam (AIP),  
Max-Planck-Institut f\"ur Astronomie (MPIA Heidelberg), 
Max-Planck-Institut f\"ur Astrophysik (MPA Garching), 
Max-Planck-Institut f\"ur Extraterrestrische Physik (MPE), 
National Astronomical Observatories of China, New Mexico State University, 
New York University, University of Notre Dame, 
Observat\'ario Nacional / MCTI, The Ohio State University, 
Pennsylvania State University, Shanghai Astronomical Observatory, 
United Kingdom Participation Group,
Universidad Nacional Aut\'onoma de M\'exico, University of Arizona, 
University of Colorado Boulder, University of Oxford, University of Portsmouth, 
University of Utah, University of Virginia, University of Washington, University of Wisconsin, 
Vanderbilt University, and Yale University.

This work has made use of data from the European Space Agency (ESA)
mission {\it Gaia} (\url{https://www.cosmos.esa.int/gaia}), processed by
the {\it Gaia} Data Processing and Analysis Consortium (DPAC,
\url{https://www.cosmos.esa.int/web/gaia/dpac/consortium}). Funding
for the DPAC has been provided by national institutions, in particular
the institutions participating in the {\it Gaia} Multilateral Agreement.

This publication makes use of data products from the Two Micron All Sky Survey, which is a joint project of the University of Massachusetts and the Infrared Processing and Analysis Center/California Institute of Technology, funded by the National Aeronautics and Space Administration and the National Science Foundation.

This research has made use of the SIMBAD database, operated at CDS, Strasbourg, France.
\end{acknowledgements}

\bibliographystyle{aa}
\bibliography{mybib.bib}




\end{document}